\documentclass[final,english]{bullsrsl}[2022/06/15]

\usepackage[latin1]{inputenc}
\usepackage[T1]{fontenc}

\usepackage{natbib} 
\usepackage{graphicx}
\usepackage{xspace}	

\newcommand{\AstroSat}{{\em AstroSat}\xspace}
\newcommand{\fermi}{{\em Fermi}\xspace}

\newcommand{\swift}{{Neil~Gehrels~\em Swift}\xspace}
\newcommand{\tninty}{{$T_{\rm 90}$}\xspace}

\begin{document}
\title{Recent observations of peculiar Gamma-ray bursts using 3.6\,m Devasthal Optical Telescope (DOT)}

\author[affil={1,2},corresponding]{Rahul}{Gupta}
\author[affil={1}]{S.}{B. Pandey}
\author[affil={1,3}]{Amit}{K. Ror}
\author[affil={1,2}]{Amar}{Aryan}
\author[affil={2}]{S.}{N. Tiwari}
\affiliation[1]{Aryabhatta Research Institute of Observational Sciences (ARIES), Nainital-263002, India}
\affiliation[2]{Department of Physics, Deen Dayal Upadhyaya Gorakhpur University, Gorakhpur-273009, India}
\affiliation[3]{Department of Applied Physics, Mahatma Jyotiba Phule Rohilkhand University, Bareilly-243006, India}
\correspondance{rahulbhu.c157@gmail.com, rahul@aries.res.in}
\date{31st May 2023}
\maketitle

\begin{abstract}

India has been actively involved in the follow-up observations of optical afterglows of gamma-ray bursts (GRBs) for more than two decades, using the country's meter-class facilities such as the 1.04\,m Sampurnanand Telescope, 1.3\,m Devasthal Fast Optical Telescope, 2.01\,m Himalayan Chandra Telescope along with many others in the country, utilizing the longitudinal advantage of the place. However, since 2016, Indian astronomers have embarked on a new era of exploration by utilizing the country's largest optical telescope, the 3.6\,m Devasthal Optical Telescope (DOT) at the Devasthal Observatory of ARIES Nainital. This unique telescope has opened up exciting opportunities for transient study. Starting from the installation itself, the DOT has been actively performing the target of opportunity (ToO) observations, leading to many interesting discoveries. Notable achievements include the contributions towards the discovery of long GRB 211211A arising from a binary merger, the discovery of the most delayed optical flare from GRB 210204A along with the very faint optical afterglow (fainter than 25 mag in g-band) of GRB 200412B. We also successfully observed the optical counterpart of the very-high-energy (VHE) detected burst GRB 201015A using DOT. Additionally, DOT has been used for follow-up observations of dark and orphan afterglows, along with the observations of host galaxies associated with peculiar GRBs. More recently, DOT's near-IR follow-up capabilities helped us to detect the first near-IR counterpart (GRB 230409B) using an Indian telescope. In this work, we summarise the recent discoveries and observations of GRBs using the 3.6\,m DOT, highlighting the significant contributions in revealing the mysteries of these cosmic transients.
\end{abstract}


\section{Introduction}

GRBs are the most energetic astronomical sources in the Universe. The initial prompt emission phase of GRBs lasts only a few seconds to minutes. The prompt phase is followed by a longer-lived/ multi-wavelength afterglow phase that can continue for days, weeks, or even months \citep{1995ARA&A..33..415F, 2013FrPhy...8..661G, 2015PhR...561....1K}. While the prompt emission can be detected by satellites like \fermi, \swift, \AstroSat, and {\it INTEGRAL}, etc., the study of afterglows necessitates the collective efforts of ground-based/space-based observatories and a global network of astronomers collaborating across the electromagnetic band. The broadband afterglow originates from the interaction between the jet and the surrounding circumburst medium. These interactions give rise to two distinct shocks known as the forward shock (FS) and the reverse shock (RS) \citep{1997ApJ...476..232M, Sari:1999}. These shocks play important roles in shaping the observed afterglow emission across different wavelengths. 
 
The FS is created when the jetted ejecta interacts with the surrounding external medium \citep{1997ApJ...476..232M, 1998ApJ...497L..17S, 2001548787S, 2000ApJ...537..255D, 2001ApJ...559..110Z}. The interaction causes the acceleration of electrons which emit the synchrotron radiation. This synchrotron emission extends from radio to X-ray wavelengths and is usually characterized by a power-law decay in the light curve. The evolution of the FS emission is dependent on the density profile of the ambient medium and other microphysical parameters. Detailed observations and modeling of the FS component provide crucial information about the energetics of the burst, the nature of the ambient medium, and the dynamics of the jet \citep{2022ApJ...940..169D}.

The FS is also accompanied by a RS component in the afterglow emission. The RS component is created when a relativistic jet interacts with slower-moving material within the ejecta itself. The reverse shock propagates back into the relativistic ejecta, and it also emits synchrotron radiation. The RS emission is usually observed in the early stages of the afterglow and is characterized by a distinct spectral and temporal behaviour \citep{2003Natur.422..284F, 2000ApJ...542..819K, 2000ApJ...545..807K, 2003ApJ...597..455K, 2014ApJ...785...84J}. RS can provide valuable insights into the properties of the ejecta, such as its magnetization and composition. 

India has been actively involved in the follow-up observations of optical afterglows of GRBs for more than two decades, using the country's meter-class facilities \citep{1999BASI...27....3S, 2000BASI...28...15S, 2000BASI...28..499S, 2001BASI...29..459P, 2003A&A...408L..21P, 2004A&A...417..919P, 2005BASI...33..209S}. In this research article, we present our efforts for the follow-up observations of the afterglow using recently installed India's largest 3.6\,m DOT telescope. We briefly summarise the recent discoveries in the field of GRBs. We investigated the temporal and spectral properties of afterglows and examined the trends that emerge from the observed data. In section \ref{DOT}, we present the afterglow observations using the DOT facility, followed by key results in section \ref{results}. We have given a summary and conclusion of the work in section \ref{summary}.

\section{GRBs follow-up observations using 3.6 m DOT}
\label{DOT}

The 3.6\,m DOT is a state-of-the-art astronomical telescope located at the Devasthal in the state of Uttarakhand, India. It is one of the largest and most advanced optical telescopes in Asia. It was commissioned in 2016, and since then, the telescope has been managed by ARIES, Nainital. Its construction was undertaken by the Belgian company AMOS (Advanced Mechanical and Optical Systems). The DOT site's high altitude (approximately 2,500 meters) and favourable climate contribute to reduced light pollution, minimal atmospheric turbulence, and clear skies, providing astronomers with optimal conditions for their research \citep{2018BSRSL..87...29K}. The telescope is equipped with four sophisticated instruments and detectors, enabling a wide range of astronomical observations across different wavelengths. These instruments include 1. 4K $\times$ 4K IMAGER: This instrument is mainly used for deep optical photometric observations and is very useful for faint objects like the afterglow of GRBs \citep{2018BSRSL..87...42P}. 2. Aries-Devasthal Faint Object Spectrograph \& Camera (ADFOSC): This versatile instrument combines a low-resolution spectrograph with an imaging camera \citep{2019arXiv190205857O}. 3. TIFR-ARIES Near Infrared Spectrometer (TANSPEC): This instrument is designed to observe (both imaging and spectroscopy capabilities) in the near-infrared portion of the electromagnetic spectrum \citep{2022PASP..134h5002S}. 4. TIFR Near Infrared Imaging Camera - II (TIRCAM2): This instrument specializes in photometric observations in near-infrared bands \citep{2018JAI.....750003B}.

When the prompt emission is detected by high-energy satellites such as \fermi \citep{2009ApJ...697.1071A, 2009ApJ...702..791M} or {\em Swift} \citep{2004ApJ...611.1005G}, rapid follow-up observations are important for studying the afterglow emission across different wavelengths. The DOT benefits from its location for transient follow-up observations as it lies nearly a 180-degree wide band between the Canary island and eastern Australia. The 3.6\,m DOT, with its advanced instrumentation and observational capabilities and longitudinal advantages, contributes significantly to these follow-up campaigns \citep{2016RMxAC..48...83P}. 

1. Afterglow Monitoring: The DOT's long-term monitoring capabilities make it well-suited for tracking and studying the evolution of GRB afterglows in the optical wavelength. This monitoring enables detailed investigations of the afterglow light curve, spectral changes, and variability, contributing to our understanding of the physical processes occurring in the aftermath of a GRB.

2. Multi-Band Photometry: The 3.6\,m DOT is equipped with a suite of filters covering a wide range of wavelengths (optical-near IR). This enables multi-filter photometric observations of the afterglows, which are crucial for constraining the spectral energy distribution (SED) and determining the spectral index, the position of a spectral break, and possible extinction in the local host. 

3. Redshift Measurements: The DOT's spectroscopic capabilities are important for measuring the redshift, which provides vital information about the distances and cosmological origins of GRBs. 

4. Host galaxies observations: The study of host galaxies of GRBs provides valuable insights into the environments of GRBs, as well as the nature of their progenitors. The 3.6\,m DOT plays an important role in observing and characterizing the host galaxies, contributing to our understanding of the broader context of these powerful events. The host galaxies' observations provide redshift measurements, star formation rate, metallicity, stellar mass, and information about other parameters of the galaxies. It also helps in examining the presence of dust extinction within the host galaxy. Such investigations help to characterize the conditions and triggering mechanisms that lead to GRB events.

\subsubsection{Observing strategy: ToO trigger method}

\begin{figure}[!t] 
\centering
\includegraphics[height=16cm,width=20.05cm,angle=90]{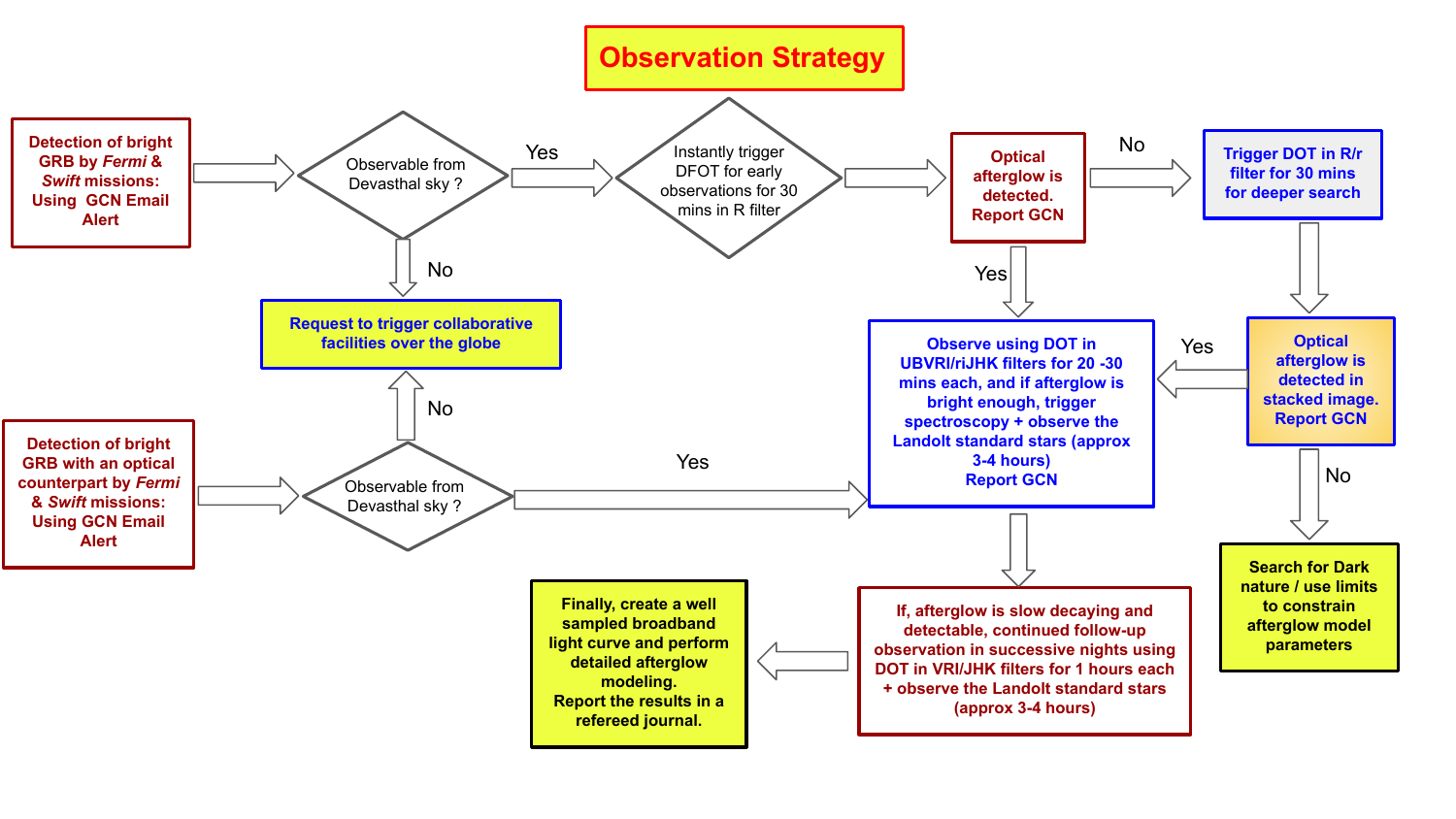}
\bigskip
\begin{minipage}{16cm}
\caption{The flow diagram for the observation plan of GRBs afterglows using 3.6\,m DOT.}
\label{fig1}
\end{minipage}
\end{figure}

For bright GRBs, we usually perform initial and quick follow-up observations using ground-based optical meter class telescopes. These facilities are used for early afterglow detection and follow-up observations. The early detection by these smaller telescopes suggests the brightness and decay rate of the afterglows. If the afterglow is bright, we trigger spectroscopic observations for the redshift measurement using DOT (If ADFOSC is available). Once the afterglow faded significantly, we generally use DOT for late-time deep photometric follow-up observations of these sources. If the afterglow is faint ($\sim$19-20 mag) during the early observations, we directly trigger the DOT for multi-band photometric observations and follow the source till the detection limit. We generally also observe the  standard stars to calibrate both photometric and spectroscopic observations. Subject to the brightness of the afterglow, we performed another observational set on successive nights. For relatively closer GRBs, late-time observations ($\sim$ 1-2 weeks after the detection) can provide a unique opportunity to detect the underlying supernova emission. In addition to the above-mentioned observational plan, the planning sometimes varies depending on the afterglow's brightness and decay rate. A brief observation strategy is also shown in Fig. \ref{fig1}. Following the above criteria, ARIES telescopes have triggered for $\sim$ 65 GRB afterglow follow-up observations (since cycle 2020C2).

\section{Results}
\label{results}
This section highlights key results obtained using 3.6\,m DOT in recent times. Brief properties of the bursts presented in this work (see references for details) are listed in Table \ref{sample_table}.

\begin{table*}
\scriptsize
\caption{The basic prompt and afterglow characterise of some of the GRBs (presented in this work) observed using 3.6\,m DOT.}
\label{sample_table}
\begin{center}
\begin{tabular}{ |c|c|c|c|c|c|c|c|c|c|}
\hline
Sr. No. & GRB name & Redshift & \tninty (sec) & Fluence ($\rm erg ~cm^{-2}$) & SN/KN & X-ray & Optical & Radio & References \\
\hline
1& GRB 211211A & 0.108 & 34.305 $\pm$ 0.572 & 5.0118 $\times$ 10$^{-4}$ & KN & Y & Y & N & \cite{2022Natur.612..228T} \\
2& GRB 210204A & 1.1710 & 206.852 $\pm$ 2.290 & 7.5698 $\times$ 10$^{-5}$ & Flare & Y & Y & Y & \cite{2022MNRAS.513.2777K} \\
3& GRB 171010A & 0.3293 & 107.266 $\pm$ 0.810 & 6.3279$\times$ 10$^{-4}$  & SN & Y& Y& Y& \cite{2022NewA...9701889K} \\
4& GRB 171205A &0.0368 &  189.4$\pm$35.0 & 3.6$\times$ 10$^{-6}$  & SN & Y&Y &Y &  \cite{2022NewA...9701889K}\\
5& GRB 201015A &0.4260 & 9.780 $\pm$ 3.47&   2$\times$ 10$^{-7}$ & SN & Y & Y& Y&  \cite{2023ApJ...942...34R}   \\
6& GRB 210205A & -- & 22.70 $\pm$ 4.18 & 8.7$\times$ 10$^{-7}$ & -- & Y& N&N & \cite{2022JApA...43...11G}\\
7& AT2021any & 2.514& -- & -- & -- & Y& Y& N& \cite{2022JApA...43...11G} \\
8& GRB 230409B& --& 9.79 $\pm$ 2.59 & 9.9$\times$ 10$^{-7}$ & -- &Y & Y&N & \cite{2023GCN.33627....1G} \\
\hline
\end{tabular}
\end{center}
\end{table*}

\subsection{GRB 211211A: A long GRB connected with Kilonova emission}

GRBs are classified based on their observed properties, duration, and spectral characteristics. Long-duration GRBs (L-GRBs) generally last longer than 2 sec and exhibit complex light curves with multiple peaks. They are often associated with the collapse of massive stars, specifically the core collapse of massive, rapidly rotating stars. On the other hand, short-duration GRBs (S-GRBs) have durations of less than 2 sec and often show simpler, single-peak light curves. A subset of L-GRBs shows evidence of associated supernova (SN) emission. S-GRBs are believed to originate from the coalescence of compact objects, such as neutron star mergers or black hole-neutron star mergers. A few short GRBs show evidence of associated Kilonovae (KN) emissions. 

A nearby GRB 211211A was discovered by \fermi and \swift missions on 11$^{th}$ December 2021 \citep{2021GCN.31201....1F, 2021GCN.31202....1D, 2021GCN.31210....1M}. The follow-up observation of GRB 211211A by DOT commenced 0.37 days after the initial burst trigger. Subsequently, we conducted nightly observations of the burst using the 4K $\times$ 4K instrument of the 3.6\,m DOT telescope. Multiple observations were captured across various optical filters  \citep{2021GCN.31299....1G}. The comprehensive collection of early to late time data acquired from the 3.6\,m DOT telescope was crucial in detecting KN emissions associated with GRB 211211A. This groundbreaking discovery signifies the first identification of a merger event as the source of a long GRB \citep{2022Natur.612..228T}.

\subsection{GRB 210204A: The detection of most delayed optical flare}

GRB 210204A was discovered by the Fermi with an error circle of 4 degrees \citep{2021GCN.29390....1F}. The Zwicky Transient Facility (ZTF) reported the discovery of a candidate afterglow of this GRB around 38 mins after the trigger \citep{2021GCN.29405....1K}. The observation of GRB 210204A was initiated by DOT $\sim$ 2.4 days following the trigger event. The 4K $\times$ 4K instruments aboard DOT were utilized, employing several optical filters \citep{2021GCN.29490....1G}. The observation of this burst by DOT extended for a continuous period of around 20 days. Notably, the results obtained from DOT's observations revealed an interesting discovery---our late-time data confirmed the detection of most delayed late optical flare associated with GRB 210204A. We extensively investigated various potential phenomena that could account for such flares, including SN emission, patchy shell emission, reverse shock ejecta, the collision of multiple forward shocks, and others. However, the most plausible explanation emerged as the refreshed shock, which caused the flare observed in the later stages of the optical light curve \citep{2022MNRAS.513.2777K}.

\subsection{GRB 171010A and GRB 171205A: Two nearby SNe-connected long GRBs powered by magnetar}

The central engine is responsible for producing an immense amount of energy. The central engine of GRB is believed to involve a compact object, such as a stellar-mass black hole or a rapidly rotating neutron star (also known as a magnetar). Two nearby bursts, GRB 171010A and GRB 171205A, were discovered by \fermi and \swift missions, respectively \citep{2017GCN.21985....1O, 2017GCN.22177....1D}. In our recent study \citep{2022NewA...9701889K}, we presented multi-band late-time follow-up observations of these two nearby supernovae-associated long GRBs. GRB 171010A was observed by the DOT starting 42 days after the burst occurred, revealing the underlying SN emission and the second burst, GRB 171205A, was observed 105 days after the burst. Notably, the detection of the underlying SN 2017htp associated with the bright GRB 171010A represents one of the most luminous GRB-SNe events. Conversely, SN 2017iuk, associated with the less bright GRB 171205A, corresponds to a low luminosity supernova. We attempted semi-analytical light-curve modeling of these sources to constrain their power mechanisms and noted that the millisecond magnetar model nicely reproduced the bolometric light curves. To further confirm the central engine of these GRBs, we utilized the early prompt and afterglow observations considering the fact that the maximum energy budget of magnetars is 10$^{52}$ erg. Our prompt emission/afterglow results also supported magnetar as a central engine powering source.

\subsection{GRB 201015A: A nearby VHE detected burst}

VHE GRBs are recently discovered by ground-based Cherenkov telescopes. The exact mechanisms responsible for producing VHE emission in GRBs are still a topic of active research. The \swift observed GRB 201015A \citep{2020GCN.28632....1D}, detected by the MAGIC at VHE energies \citep{2020GCN.28659....1B}, represents a significant milestone as it is the first of its kind to be observed by the DOT, starting only 0.5 days after the event. The comprehensive analysis of this burst, including observations in multiple bands, was presented in the study by \cite{2023ApJ...942...34R}. A notable feature of this burst is the rarely observed bump in the early optical light curve, which plays a crucial role in determining the physical properties of the ultrarelativistic jet. These properties include the composition of the jet and the Lorenz factor of the burst. Additionally, the late-time follow-up observations are equally important as they provide insights into the spectral regime and the surrounding environment of the burst. The DOT's observations of this burst have proven invaluable in constraining the optical spectral index ($\beta$) during the later stages of the event, which, in turn, has helped constrain the ISM-like medium surrounding the burst \citep{2023ApJ...942...34R}.

\subsection{GRB 210205A and AT2021any: Dark and orphan GRBs}

A significant portion of GRBs either lack detectable optical afterglows or exhibit very faint signals, even after extensive and deep searches using optical telescopes. These particular afterglows are commonly referred to as Dark GRBs. Various factors can contribute to the optical darkness of GRB afterglows, such as intrinsic faintness, high redshift, and obscuration scenario. The \swift X-ray Telescope (XRT) identified an X-ray afterglow associated with GRB 210205A \citep{2021GCN.29397....1D}. Subsequently, we conducted optical follow-up observations using the 3.6\,m DOT to search for the optical counterpart \citep{2021GCN.29526....1P}. Despite extensive and deep observations, we were unable to detect any optical afterglow, thus confirming the dark nature of this burst. In our investigation of the potential causes behind this optical darkness, we find two possibilities: intrinsic faintness of the source or high redshift origin \citep{2022JApA...43...11G}.

There exists a small number of rare cases where optical afterglows are detected independently without any preceding detection of prompt emission from space-based gamma-ray satellites. These peculiar instances are referred to as ``Orphan afterglows." Several possible explanations have been proposed to account for the occurrence of orphan afterglows: limited sensitivity, Off-axis Observations, and absorption of gamma rays. We studied an orphan case AT2021any for which no gamma-ray prompt emission counterpart was reported by any space-based satellites. Surprisingly, this source was independently discovered by the ZTF \citep{2021GCN.29305....1H}. Subsequently, we detected and studied this object using the 3.6\,m DOT. This discovery is noteworthy as AT2021any represents the third known orphan afterglow for which a redshift measurement has been obtained. To gain insights into the nature of AT2021any, we conducted a comprehensive analysis involving detailed broadband afterglow modeling. Our investigations confirmed that AT2021any was observed on-axis, and the gamma-ray counterpart of AT2021any was missed by GRB satellites \citep{2022JApA...43...11G}.

\subsection{GRB 230409B: The first near-IR afterglow detected using Indian telescope}

\begin{figure}[!t] 
\centering
\includegraphics[height=8.5cm,width=9cm,angle=0]{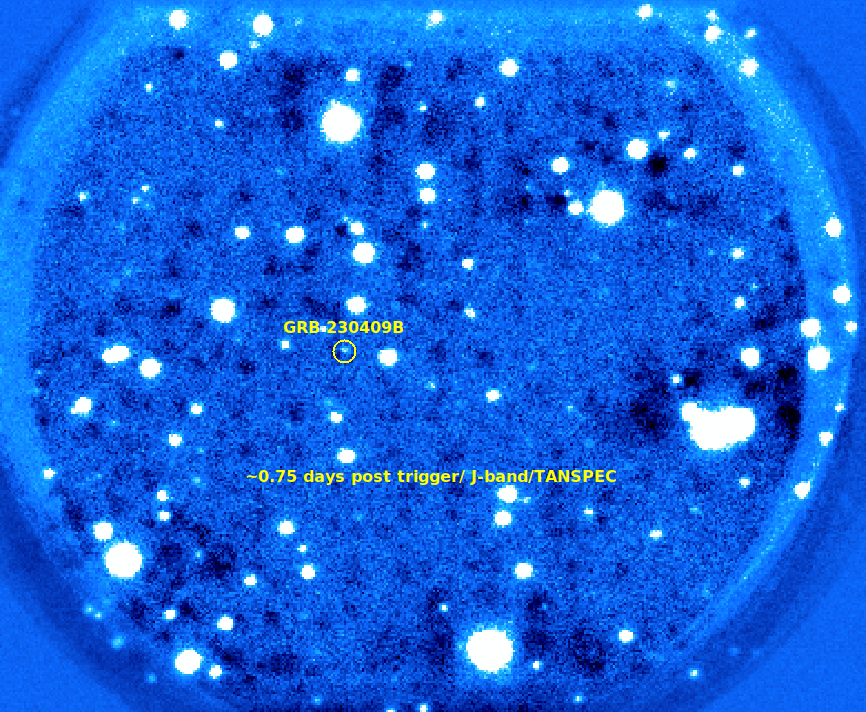}
\bigskip
\begin{minipage}{16cm}
\caption{Detection and confirmation of the near-IR ($J$-band) afterglow of GRB 230409B taken using 3.6\,m DOT. The afterglow position is marked using a yellow circle. It is the first case of near-IR afterglow being discovered using any Indian telescope.}
\label{fig2}
\end{minipage}
\end{figure}

GRB 230409B was detected by the \swift Burst Alert Telescope (BAT) on April 9, 2023. The XRT detected an X-ray afterglow approximately 101.2 seconds after the BAT trigger \citep{2023GCN.33592....1B}. The \swift Ultraviolet/Optical Telescope (UVOT) detected a weak fading source within the XRT error circle \citep{2023GCN.33595....1K}. To further investigate this fading optical source, we conducted follow-up observations using the TANSPEC instrument of the 3.6\,m DOT. Multiple observations were carried out in the J filter (near-infrared) at different epochs. The brightness of the afterglow was determined to be J = 20.1 $\pm$ 0.3 magnitude approximately 0.75 days after the BAT trigger \citep{2023GCN.33627....1G}. This is the first near-IR afterglow detected using an Indian telescope (see Fig. \ref{fig2}). Subsequent observations yielded a limiting magnitude of 20.5 mag around 2.7 days after the burst. Our observations conclusively confirmed that this fading source corresponds to the afterglow of GRB 230409B \citep{2023GCN.33627....1G}.

\subsection{Host galaxies of GRBs}

We have also directed our attention to the host galaxies associated with select intriguing GRBs using the DOT. The study of GRB host galaxies presents a unique opportunity to estimate essential properties such as stellar mass, ages, star-formation rates, and other vital characteristics of the burst environments and their progenitors. We employed a detailed SED modeling approach utilizing an advanced tool known as Prospector. Furthermore, we compared the results obtained from our SED modeling with a larger sample comprising well-studied host galaxies of GRBs, supernovae, and normal star-forming galaxies \citep{2022JApA...43...82G}.

\section{Summary and Conclusion}
\label{summary}

The afterglow observations provide crucial insights into the dynamics, energetics, and jet composition, as well as the environments in which these events occur. In this work, we present the recent key results (see below) in the research field of GRBs obtained using 3.6\,m DOT. 

\begin{enumerate}
    \item The discovery of GRB 211211A, accompanied by kilonova emission, provides compelling evidence that certain long-duration GRBs may arise from the merger of neutron stars.
    
\item Among GRBs, GRB 210204A stands out with the most pronounced delayed flaring activity ever observed.
    
\item Semi-analytical light-curve modeling of GRB 171010A/SN 2017htp and GRB 171205A/SN 2017iuk indicates the presence of a spin-down millisecond magnetar as the central engine driving these events.
    
    \item We studied VHE-detected GRB 201015A and investigated the evolution of the bulk Lorentz factor, which offers a potential solution to understanding the jet composition of GRBs.
    
    \item Our comprehensive observations and subsequent analysis indicate that GRB 210205A exhibits a dark afterglow, while AT2021any represents an orphan GRB, lacking a prior detection of prompt emission.
    
    \item GRB 230409B is the first case of near-IR afterglow being discovered using any Indian telescope.

    \item Also, our results reveal the remarkable capabilities of the 3.6\,m DOT and its back-end instruments for conducting in-depth photometric studies of the host galaxies of energetic transients such as GRBs, supernovae, and other astrophysical sources.

\end{enumerate}

The GRB follow-up observations using DOT show an essential contribution to the global telescope network for studying these transient and energetic sources. Through its observations, the 3.6\,m DOT help in advancing our knowledge of GRBs, supporting ongoing research efforts, and facilitating collaborations with other observatories and space missions.

\begin{acknowledgments}
We thank the referee for giving positive remarks. RG and SBP acknowledge the financial support of ISRO under AstroSat archival Data utilization program (DS$\_$2B-13013(2)/1/2021-Sec.2). AA acknowledges funds and assistance provided by the Council of Scientific \& Industrial Research (CSIR), India with file no. 09/948(0003)/2020-EMR-I. This research is based on observations obtained at the 3.6\,m Devasthal Optical Telescope (DOT), which is a National Facility run and managed by Aryabhatta Research Institute of Observational Sciences (ARIES), an autonomous Institute under the Department of Science and Technology, Government of India. RG, SBP, AKR, and AA also acknowledge the Belgo-Indian Network for Astronomy and Astrophysics (BINA) project under which local support was provided during the $3^{rd}$ BINA workshop.
\end{acknowledgments}

\begin{furtherinformation}

\begin{orcids}
\orcid{0000-0003-4905-7801}{Rahul}{Gupta}
\orcid{0000-0003-3164-8056}{Amit}{Kumar Ror}
\orcid{0000-0002-9928-0369}{Amar}{Aryan}

\end{orcids}

\begin{authorcontributions}
RG and SBP started the project. AKR helps in writing the result section of the manuscript. AKR and AA contributed to observations using 3.6\,m DOT. All the authors edit the manuscript.
\end{authorcontributions}

\begin{conflictsofinterest}
The authors declare no conflict of interest.
\end{conflictsofinterest}

\end{furtherinformation}

\bibliographystyle{bullsrsl-en}

\bibliography{extra}

\end{document}